\documentclass[11pt]{article}
\usepackage[normalem]{ulem}
\pdfoutput=1 % if your are submitting a pdflatex (i.e. if you have

\usepackage{jheppub} % for details on the use of the package, please
\usepackage{url}
\usepackage{caption}
\usepackage{subcaption}
\usepackage{extarrows}

\usepackage[latin9]{inputenc}
\usepackage{float}
\usepackage{amsmath}
\usepackage{amssymb}
\usepackage{graphicx}
\usepackage{esint}
\usepackage{hyperref}
\usepackage{comment}
\usepackage{color}
\usepackage{microtype}
\usepackage{cleveref}
\usepackage{breakurl}
\usepackage{bbm}
\newcommand{\be}{\begin{equation}}
\newcommand{\ee}{\end{equation}}
\newcommand{\ben}{\begin{displaymath}}
\newcommand{\een}{\end{displaymath}}
\newcommand{\bea}{\begin{eqnarray}}
\newcommand{\eea}{\end{eqnarray}}
\def\K{K{\"a}hler }
   \newcommand{\rf}[1]{(\ref{#1})}
\newcommand{\vp}{\varphi}

\def\be{\begin{equation}}
\def\ee{\end{equation}}
\def\bea{\begin{eqnarray}}
\def\eea{\end{eqnarray}}
\def\ba{\begin{array}}
\def\ea{\end{array}}
\def\bit{\begin{itemize}}
\def\eit{\end{itemize}}

\def\vp{\varphi}

\def\rrange{0.014^{+0.010}_{-0.011}}
\def\rul{0.036}

 \makeatletter

\allowdisplaybreaks

\makeatother

\makeatletter
\DeclareRobustCommand{\rcite}[1]{%
  \rcite@aux#1,\@nil{#1}%
}
\def\rcite@aux#1,#2\@nil#3{%
  \if\relax#2\relax
    Ref.~\cite{#3}%
  \else
    Refs.~\cite{#3}%
  \fi
}
\makeatother

\hypersetup{
    colorlinks = true,
    citecolor = {blue},
    linkcolor = {blue},
    urlcolor = {blue},
}

 \title{\rm { \LARGE \bf   BICEP/Keck and Cosmological Attractors }}

\author{Renata Kallosh and }
\author{Andrei Linde}

\affiliation{Stanford Institute for Theoretical Physics and Department of Physics,\\ Stanford University, Stanford, CA 94305, USA}
\emailAdd{kallosh@stanford.edu}
\emailAdd{alinde@stanford.edu}

\abstract{We discuss implications of the latest BICEP/Keck data release for  inflationary models, with special emphasis on the cosmological attractors which can describe all presently available inflation-related observational data.   These models are compatible with any  value of the tensor to scalar ratio $r$, all the way down to $r = 0$. Some of the string theory motivated models of this class predict  $10^{{-3}} \leq r \leq 10^{{-2}}$. The upper part of this range can be explored by the ongoing BICEP/Keck observations.}

\begin{document}

\maketitle

% \tableofcontents{}

%\newpage
\section{Introduction}

\begin{figure}[H]
\vskip -5pt
\centering
\includegraphics[scale=0.8]{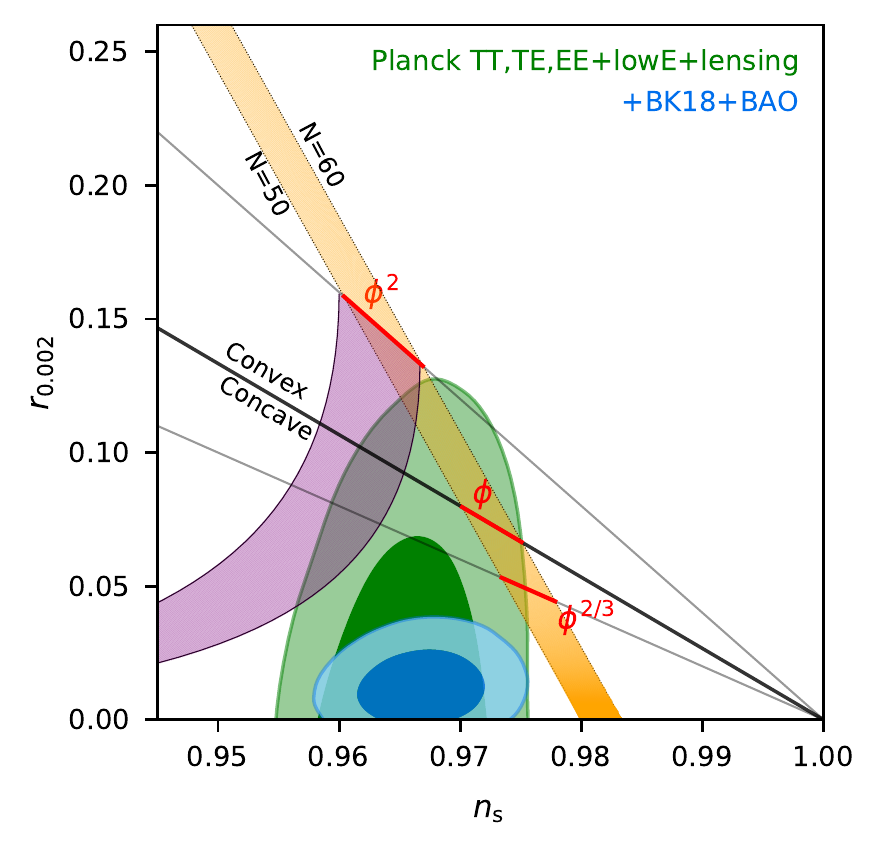}
\vskip -15pt
\caption{\footnotesize  BICEP/Keck results for $n_{s}$ and $r$ \cite{BICEPKeck:2021gln}. The $1\sigma$  and $2\sigma$  areas are represented by dark blue and light blue colors. The  purple region shows natural inflation, and the orange band corresponds to inflation driven by scalar field with canonical kinetic terms and monomial potentials.}
\label{BICEPKeck0}
\end{figure}

The new data release from BICEP/Keck considerably strengthened  bounds on the tensor to scalar ratio $r$  \cite{BICEPKeck:2021gln}: $r_{0.05}=\rrange$ ($r_{0.05}<\rul$ at 95\% confidence). The main results are illustrated in  \cite{BICEPKeck:2021gln} by a figure describing combined constraints on $n_{s}$ and $r$, which we reproduce here in Fig. \ref{BICEPKeck0}. These new results have important  implications for the development of inflationary cosmology.  In particular, the standard version of natural inflation \cite{Freese:1990rb}, as well as the full class of monomial potentials $V \sim \phi^{n} $, are now strongly disfavored.  

Additional information can be obtained for the  hilltop models. The simplest models $V = V_{0}(1-\phi^{4}/m^{4})$   represented by the green band in Fig.~8 of the Planck2018 data release  \cite{Planck:2018jri} lead to a universal prediction $n_{s} = 1-3/N_{e}$  for all sub-Planckian values of the mass parameter $m\lesssim 1$. This prediction is strongly disfavored by the Planck2018 data for the number of e-foldings $N_{e} \sim 50 - 60$.  These models could provide a good match to the Planck data for $m\gtrsim 10$. However, in that  case they predict post-inflationary collapse of the universe, which cannot be avoided without a substantial modification of such models, strongly modifying their predictions \cite{Kallosh:2019jnl}.

\begin{figure}[H]
\centering
\includegraphics[scale=0.35]{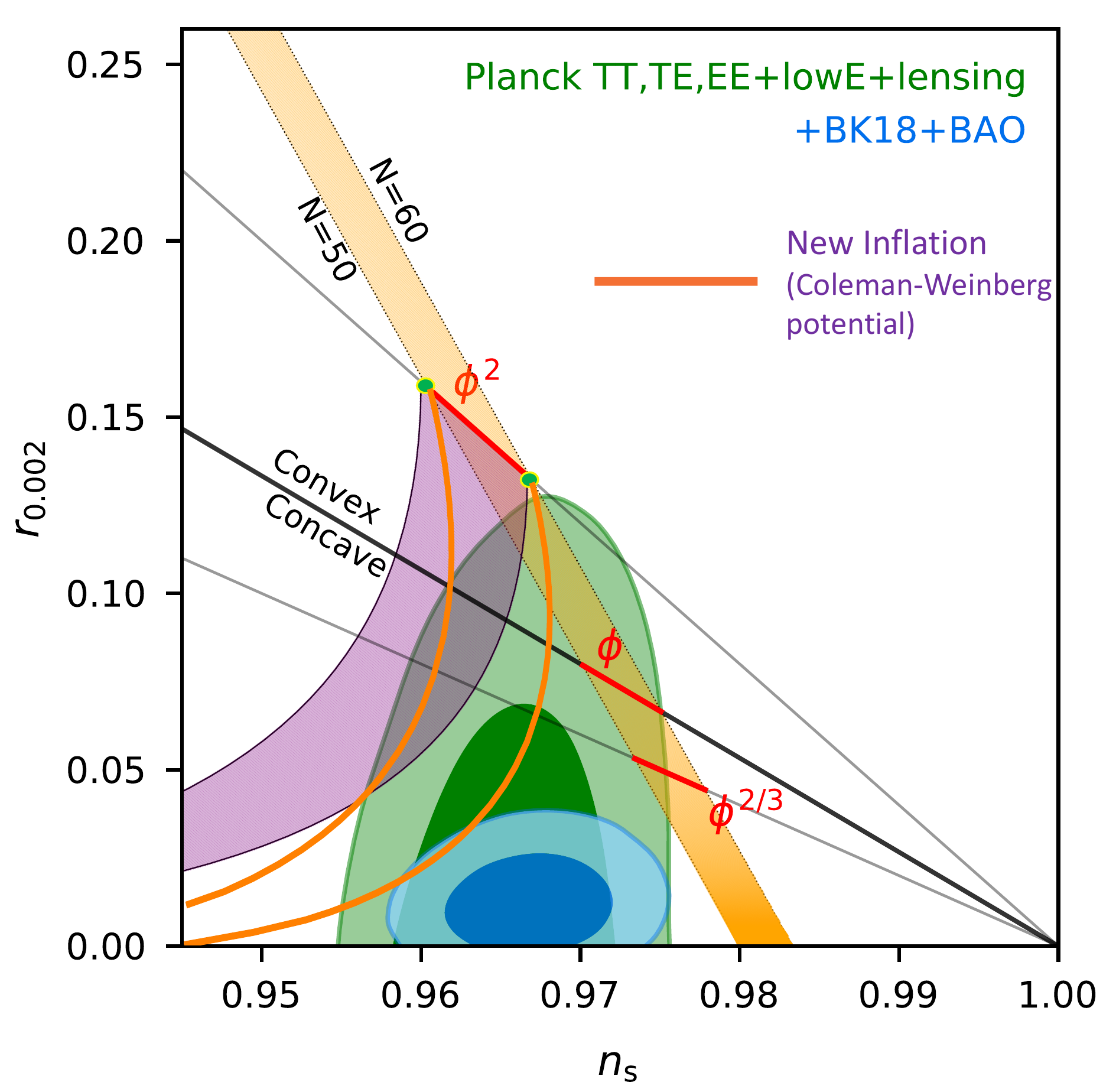}
\vskip -10pt
\caption{\footnotesize   Models of the type of new inflation \cite{Linde:1981mu,Albrecht:1982wi}  based on the Coleman-Weinberg hilltop potential  are marginally compatible with Planck2018 data, but strongly disfavored by the BICEP/Keck data  \cite{BICEPKeck:2021gln}.}
\label{NEW}
\end{figure}

More complicated versions of the hilltop models, such as the new inflation model with the Coleman-Weinberg potential $V\sim 1+{\phi^{4}\over m^{4}}(2 \log {\phi^{2}\over m^{2}}-1)$, are marginally compatible with the Planck2018 data  \cite{Kallosh:2019jnl}, though only for $m \gg 1$. Now they are strongly disfavored by the results of the recent BICEP/Keck data release, as we show in Fig. \ref{NEW}.
However, one can recover  these losses by making a  relatively simple generalization of the kinetic term of the scalar field. After this generalization, most of the improved models, which we called ``cosmological attractors,'' become compatible with all presently available inflation-related observational data, almost independently of the choice of the scalar potential prior to the generalization. 

%\newpage 
\section{\boldmath{$\alpha$-attractors}}\label{sattr}
\subsection{T-models}
We will begin with describing $\alpha$-attractors \cite{Kallosh:2013hoa,Ferrara:2013rsa,Kallosh:2013yoa,Galante:2014ifa,Kallosh:2015zsa,Kallosh:2019eeu,Kallosh:2019hzo}.
The simplest example is given by the theory
 \be
{ {\cal L} \over \sqrt{-g}} =  {R\over 2}  -  {(\partial_{\mu} \phi)^2\over 2\bigl(1-{\phi^{2}\over 6\alpha}\bigr)^{2}} - V(\phi)   \,  .
\label{cosmoA}\ee
Here $\phi(x)$ is the scalar field, the inflaton.  In the limit $\alpha \to \infty$ the kinetic term becomes the standard canonical term $-  {(\partial_{\mu} \phi)^2\over 2}$.  The new kinetic term has a singularity at $|\phi| = \sqrt{6\alpha} $. However, one can get rid of the singularity and recover the canonical normalization  by solving the equation ${\partial \phi\over 1-{\phi^{2}\over 6\alpha}} = \partial\vp$, which yields $
\phi = \sqrt {6 \alpha}\, \tanh{\varphi\over\sqrt {6 \alpha}}$.
The full theory, in terms of the canonical variables, becomes a theory with a plateau potential
 \be
{ {\cal L} \over \sqrt{-g}} =  {R\over 2}  -  {(\partial_{\mu}\varphi)^{2} \over 2}  - V\big(\sqrt {6 \alpha}\, \tanh{\varphi\over\sqrt {6 \alpha}}\big)   \,  .
\label{cosmoqq}\ee
We called such models T-models due  to their dependence on the $\tanh{\varphi\over\sqrt {6 \alpha}}$.  Asymptotic value of the potential at the plateau at large $\varphi>0$ is given by
\be\label{plateau}
V(\vp) = V_{0} - 2  \sqrt{6\alpha}\,V'_{0} \ e^{-\sqrt{2\over 3\alpha} \varphi } \ .
\ee
Here $V_0 = V(\phi)|_{\phi =  \sqrt {6 \alpha}}$ is the height of the plateau potential, and $V'_{0} = \partial_{\phi}V |_{\phi = \sqrt {6 \alpha}}$. The coefficient $2  \sqrt{6\alpha}\,V'_{0}$ in front of the exponent  can be absorbed into a redefinition (shift) of the field $\varphi$. Therefore all inflationary predictions of this theory in the regime with $e^{-\sqrt{2\over 3\alpha} \varphi } \ll 1$ are determined only by two parameters, $V_{0}$ and $\alpha$, i.e. they do not depend on any other features of the potential $V(\phi)$. That is why they are called attractors.
\begin{figure}[H]
\centering
\includegraphics[scale=0.29]{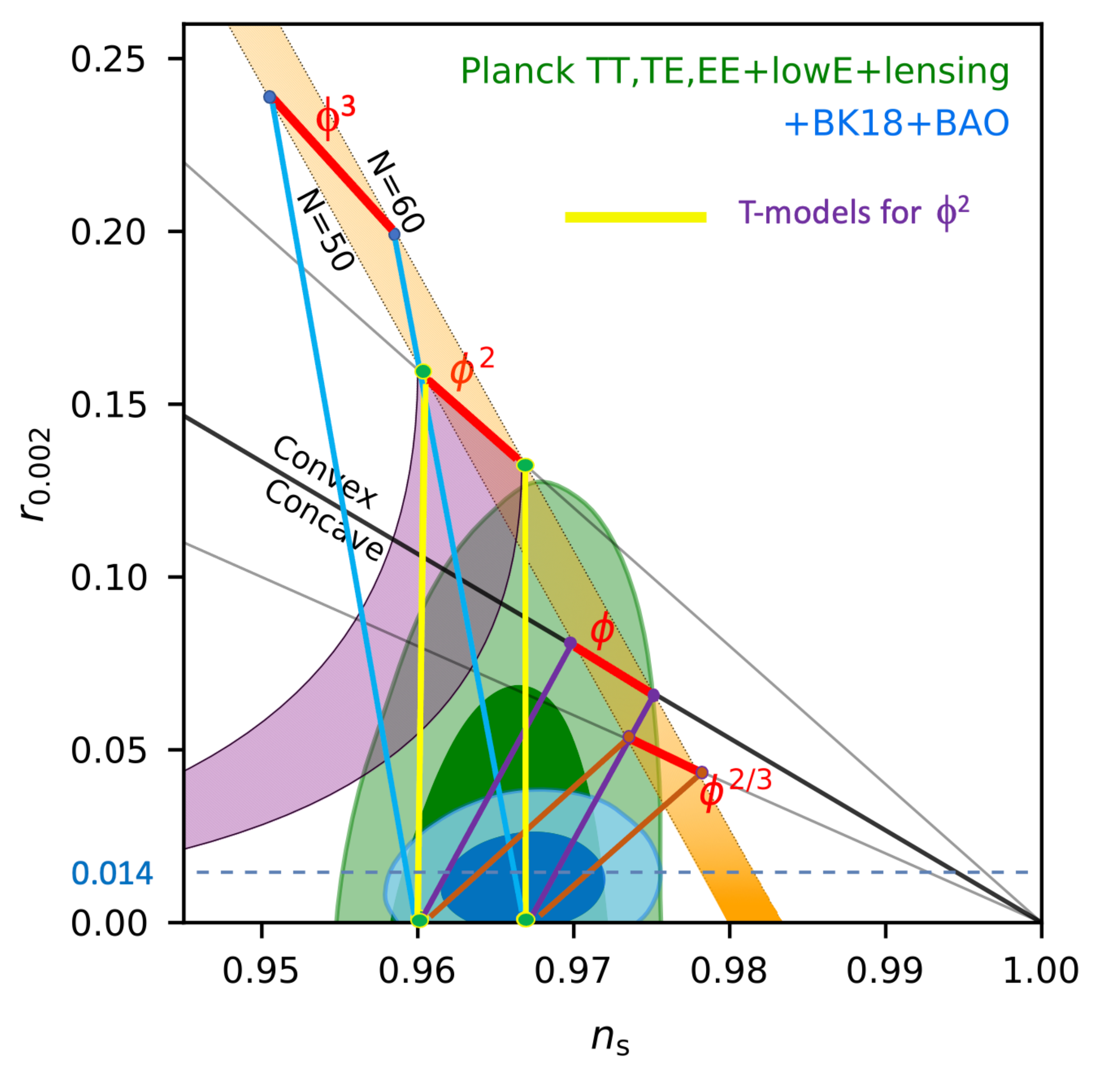}
\vskip -10pt
\caption{\footnotesize The figure illustrating the main results of the BICEP/Keck \cite{BICEPKeck:2021gln} superimposed with the predictions of  $\alpha$-attractor T-models with the potential $\tanh^{2n} {\varphi\over \sqrt{6\alpha}}$ \cite{Kallosh:2013yoa,Kallosh:2015zsa}. Each of these models starts at some $\phi^{2n}$ (at $\alpha \rightarrow \infty$)  and is forced to go down with decreasing $\alpha$ \cite{Kallosh:2013yoa} into the area favored by the BICEP/Keck. 
}
\label{Fan}
\end{figure}

To illustrate advantages of this class of models, we show in Fig. \ref{Fan} predictions of the models with monomial potentials $\phi^{2n}$ after the modification of the kinetic term shown in \rf{cosmoA}. At large $\alpha$, predictions of all of these models coincide with the predictions shown in Fig. \ref{BICEPKeck0}, and these models are ruled out, but at smaller $\alpha$ they all run towards the dark blue area favored by the latest BICEP/Keck data release.  Fig. \ref{Fan} illustrates the main advantage of the cosmological attractors: At large $N_{e}$, their predictions for $A_{s}$, $n_{s}$ and $r$ coincide in the small $\alpha$ limit, nearly independently of the detailed choice of the potential $V(\phi)$:
\be
\label{pred}
 A_{s} = {V_{0}\, N_{e}^{2}\over 18 \pi^{2 }\alpha} \ , \qquad n_{s} = 1-{2\over N_{e}} \ , \qquad r = {12\alpha\over N^{2}_{e}} \ .
\ee 
These models are compatible with the presently available observational data for sufficiently small $\alpha$.

Importantly, these results depend on the height of the inflationary plateau, which is given by $V_0 = V(\phi)|_{\phi =  \sqrt {6 \alpha}}$, but they do not depend on any other details of behavior of the  potential $V(\phi)$ in \rf{cosmoA}. This explains, in particular, stability of the predictions of these models with respect to quantum corrections \cite{Kallosh:2016gqp}.

The amplitude of inflationary perturbations  in these models matches the Planck normalization $A_{s} \approx 2.01 \times 10^{{-9}}$ for  $ {V_{0}\over  \alpha} \sim 10^{{-10}}$, $N_{e} = 60$, or for $ {V_{0}\over  \alpha} \sim 1.5 \times 10^{{-10}}$, $N_{e} = 50$.  For the simplest  model $V = {m^{2}\over 2} \phi^{2}$  one finds
\be\label{T}
V =  3m^{2 }\alpha \tanh^{2}{\varphi\over\sqrt {6 \alpha}} \ .
\ee
This simplest model  is shown by the prominent vertical yellow band in Fig. 8 of the paper on inflation in the Planck2018 data release  \cite{Planck:2018jri}.  In this model,  the condition $ {V_{0}\over  \alpha} \sim 10^{{-10}}$ reads $ m  \sim   0.6 \times10^{{-5}}$. The small magnitude of this parameter  accounts for the small amplitude of perturbations $A_{s} \approx 2.01 \times 10^{{-9}}$. No other parameters are required to describe all presently available inflation-related data in this model. If the inflationary gravitational waves are discovered, their amplitude can be accounted for by the  choice of the parameter $\alpha$ in \rf{pred}.

\subsection{E-models}
The second family of $\alpha$-attractors  called E-models is given by
 \be\label{actionE}
{ {\cal L} \over \sqrt{-g}} =  {R\over 2} - {3\alpha\over 4} \, {(\partial \rho)^2\over \rho^{2}}- V(\rho) \ .
\ee
As before, one can go to canonical variables, $\rho = e^{-\sqrt {2 \over 3\alpha}\varphi}$, which yields
\be\label{apole}
{ {\cal L} \over \sqrt{-g}} =  {R\over 2} - {1\over 2}  (\partial \varphi)^2- V(e^{-\sqrt {2\over 3\alpha}\varphi}). 
\ee
We consider $V(\rho)$  not singular at $\rho = 0$, e.g. $V(\rho) =V_{0}(1-\rho)^{2}$. In  canonical variables it gives
\be\label{E}
 V = V_0 \Bigl(1 - e^{-\sqrt {2 \over 3\alpha}\varphi} \Bigr)^{2} . 
\ee
For the particular case $\alpha = 1$  this potential coincides with the potential of the Starobinsky model  \cite{Starobinsky:1980te}. In the small $\alpha$   limit the predictions of the E-models coincide with the predictions of the T-models \rf{pred}.

\begin{figure}[H]
\centering
\includegraphics[scale=0.36]{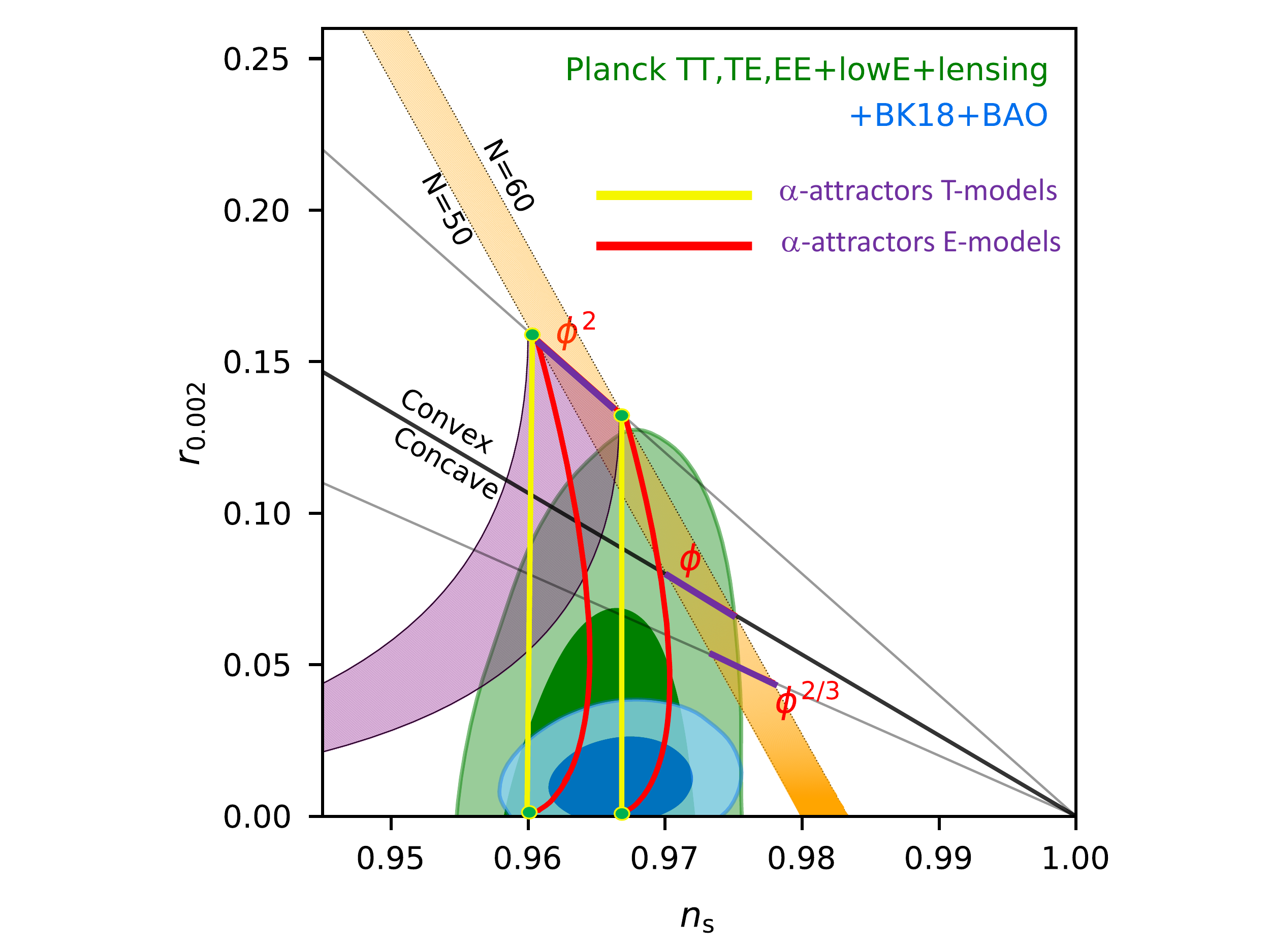}
\vskip -10pt
\caption{\footnotesize The  BICEP/Keck \cite{BICEPKeck:2021gln} figure superimposed with the predictions of the simplest $\alpha$-attractor T-model with the potential $\tanh^{2} {\varphi\over \sqrt{6\alpha}}$ and E-models (yellow lines for $N_{e} = 50, 60$) with the potential $\big (1-e^{-\sqrt{{2\over 3\alpha}}\varphi}\big )^{2}$ (red lines for $N_{e} = 50, 60$).
}
\label{TandE}
\end{figure}

Fig. \ref{TandE} shows a combination of predictions of the simplest T-model \rf{T} and the simplest E-model \rf{E}.  Predictions of both of these models at large $\alpha$ coincide with the predictions of the model  $\phi^{2}$,  and then  go down into the blue area with decreasing $\alpha$. T-model band goes straight, E-model band first slightly bends to the right, to larger values of $n_s$, but later reaches the same attractor value as in  the T-model. Their predictions are consistent with the Planck/BICEP/Keck bound  $r<0.036$ for $\alpha \lesssim 7$.
Note that both models can describe {\it any}  value of $r\ll 1$, all the way down to the ultimate attractor point $r = 0$.  

\section{Other examples of cosmological attractors}
\subsection{Pole inflation, D-brane inflation}

$\alpha$-attractors represent a special version of a more general class of attractors,  the so-called pole inflation models \cite{Galante:2014ifa}. It is obtained by slightly generalizing equation \rf{actionE}:
\be
{ {\cal L} \over \sqrt{-g}} =  {R\over 2} - {a_q\over 2} {(\partial \rho)^2 \over \rho^{q}} - V(\rho) \ .
\label{actionQ}\ee
Here the pole of order $q$ is at $\rho=0$ and  the residue at the pole is $a_q$. For $q = 2$, $a_2 = {3\alpha \over 2}$, this equation describes E-models of $\alpha$-attractors, but here we   consider general values of $q$. For $q \not = 2$ one can always rescale $\rho $ to make $a_q=1$.  Just as in the theory of $\alpha$-attractors, one can make a transformation to the canonical variables $\varphi$ and find that the asymptotic behavior of the potential $V(\varphi)$ during inflation is determined only by  $V(0)$ and the first derivative ${dV(\rho)\over d\rho}|_{\rho = 0}$ . The value of $n_{s}$ for this family of attractors  is given by 
\be
n_s= 1 - {\beta\over N_e}\, , \qquad \beta = {q \over q-1} \ .
\label{ns}
\ee
We will discuss here the models with $q > 2$, $\beta < 2$, which provide  spectral index $n_{s}$ slightly greater than  the $\alpha$-attractors result $n_s= 1 - {2\over N_e}$ . 

For $\alpha$-attractors  the plateau of the potential  is reached exponentially. For $q > 2$ the approach to the plateau is controlled by  negative powers of $\varphi$. Some of these models described in  \cite{Kallosh:2018zsi,Kallosh:2019eeu,Kallosh:2019hzo}   have interpretation in terms of Dp-brane inflation \cite{Dvali:1998pa,Kachru:2003sx}. 
The ${\rm Dp}-{\overline {\rm Dp}}$  brane inflation plateau potentials are
\be
V_{Dp-\bar Dp} \sim   {|\varphi|^k\over m^k + |\varphi|^k}= \Big( 1+  {m^k\over |\varphi|^k }\Big )^{-1}\, , \qquad k=7- p= {2\over 2-q} \ .
\label{KKLTI}\ee
Their attractor formula for the $n_s$ is given in \rf{ns}, whereas the formula for $r$ depends on the parameter $m$ in the potential.
For ${\rm D3}-{\overline {\rm D3}}$ inflation   for small $m$  one has   
\be
V = V_{0} \ {\varphi^{4}\over m^{4} + \varphi^{4}}\ , \qquad n_s =  1- {5 \over  3 N_{e}} \ , \qquad r = {4 m^{{4\over 3} }\over (3N_{e})^{5\over 3}} \ .
\ee
For  ${\rm D5}-{\overline {\rm D5}}$ brane inflation one has 
\be
V = V_{0} \ {\varphi^{2}\over m^{2} + \varphi^{2}}\ , \qquad n_s =  1- {3 \over  2 N_{e}} \ , \qquad   r=   {\sqrt 2\, m \over N_{e}^{3\over 2}} \ .
\ee
In Fig. \rf{Blue} we give a combined plot of the predictions of the simplest $\alpha$-attractor models and Dp-brane inflation for $n_{s}$ and $\log_{10}r$,  for $N_{e} = 50$ and $60$.  \cite{Kallosh:2019hzo}.  In the small $m$ limit, the predicted values of $r$ for Dp-brane inflation, and for pole inflation in general, can take extremely small values, all the way down to $r \to 0$. 
\begin{figure}[H]
\centering
\includegraphics[scale=0.31]{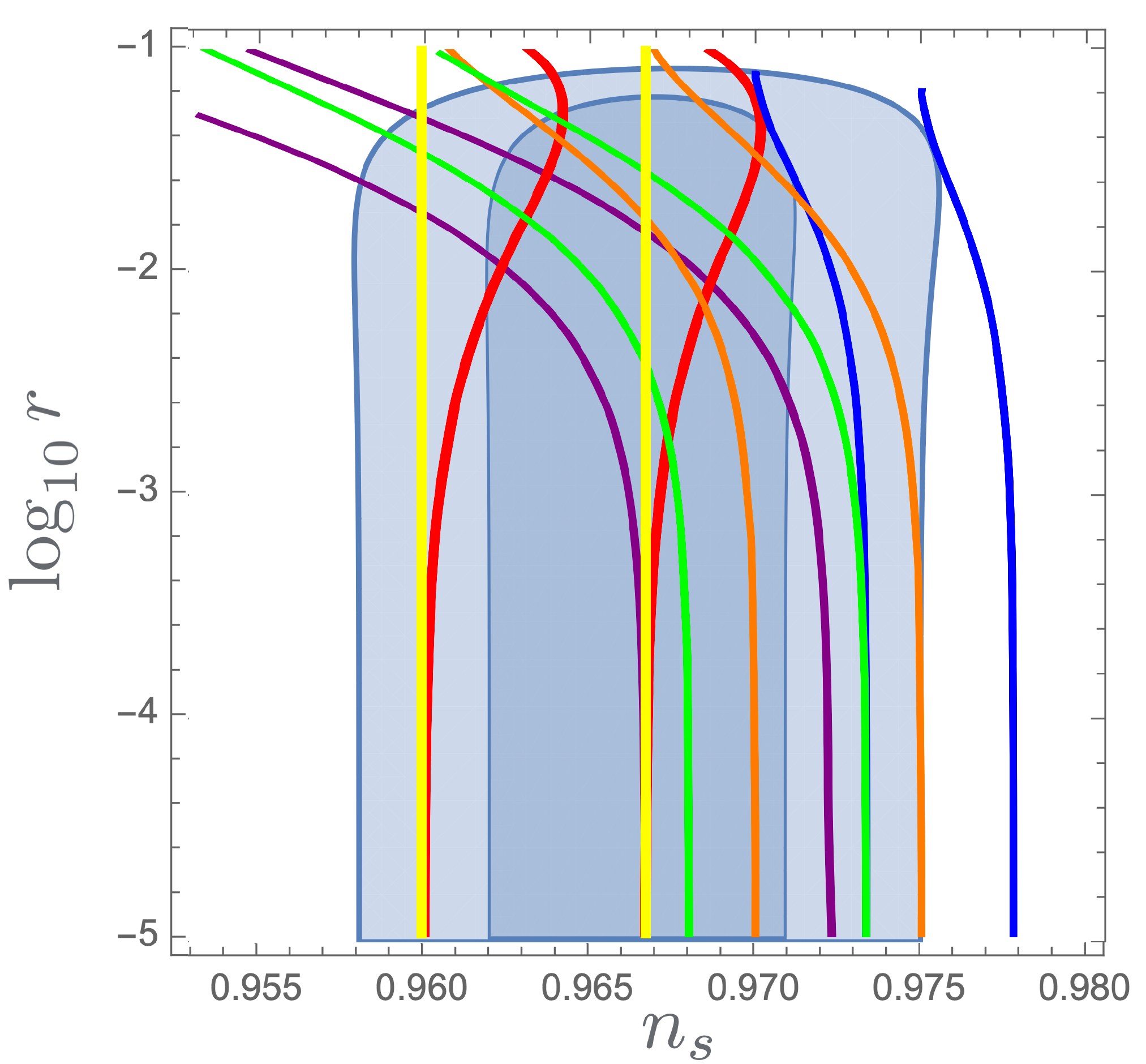}
\vskip -5pt
\caption{\footnotesize A combined plot of the predictions of the simplest $\alpha$-attractor models and Dp-brane inflation  for $N_{e} = 50$ and $60$ \cite{Kallosh:2019hzo}. From left to right, we show predictions of T-models and E-models (yellow and red lines) and of ${\rm Dp}-{\overline {\rm Dp}}$  brane inflation with $p = 3, 4, 5, 6$ (purple, green, orange and blue lines) for potentials in eq. \rf{KKLTI} with $k=4,3,2,1$. The blue data background  corresponds to  Planck 2018 results  including BAO.}
\label{Blue}
\end{figure}
The  potentials which appear in the pole inflation scenario may have an alternative interpretation, not related to Dp-branes.   For example, a quadratic model $V  \sim   {\varphi^2\over m^2 + \varphi^2}$ was proposed in \cite{Dong:2010in} as an example of a flattening mechanism for the $\varphi^2$ potential due to the inflaton interactions with heavy scalar fields. Similar potentials with flattening may also appear in axion theories in the strong coupling regime \cite{DAmico:2017cda}.

Independently of their interpretation, the  pole inflation models may serve as a powerful tool for parametrization of all observational data since all data for $n_{s}$ and $r$ can be sorted out using vertical $\beta$ stripes with $n_s= 1 - {\beta\over N_e}$ \cite{Kallosh:2019eeu,Kallosh:2019hzo}. As illustrated by Fig. \rf{Blue}, just a few of such stripes may completely cover all possible values of $n_{s}$ and $r$ compatible with the observational data. This parametrization works especially well in the small $r$ limit, which is the top priority for parametrizing the results of the ongoing and planned  search for the inflationary gravitational waves.

\subsection{$\xi$-atttractors}

\begin{figure}[H]
\centering
\includegraphics[scale=0.41]{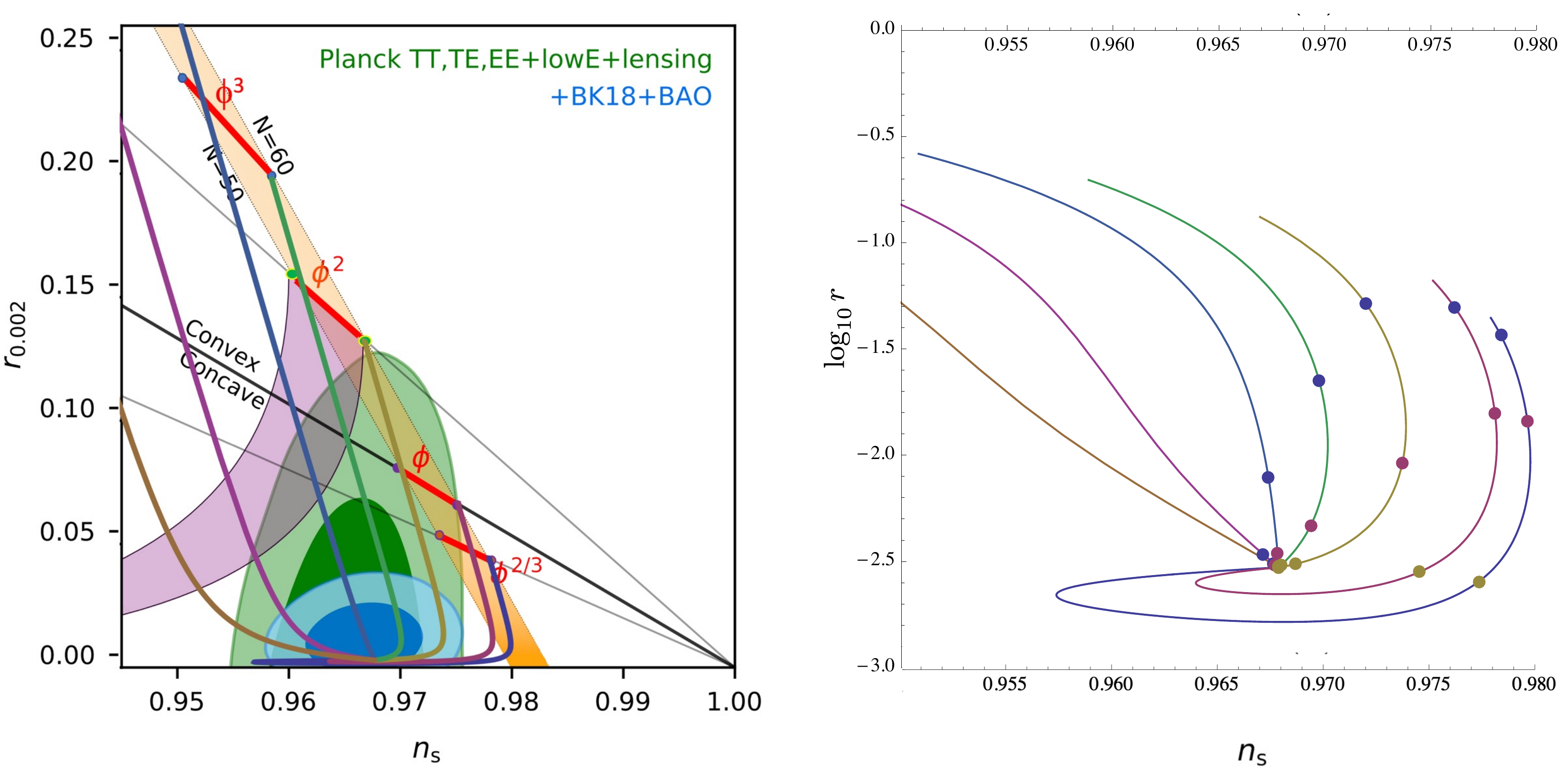}\vskip -5pt
\caption{\footnotesize  Attractor trajectories 
%on a linear and a logarithmic scale 
for  monomial models $V(\phi) \sim \phi^{8}, \phi^{6}, \phi^4, \phi^2, \phi, \phi^{2/3}$, for $N_{e}= 60$. In the left panel, the results of  \cite{Kallosh:2013tua} are superimposed with the BICEP/Keck results represented by Fig. \ref{BICEPKeck0}. The right panel shows the same attractor trajectories, but with the vertical axis corresponding to  $\log_{10}r$. This  more clearly shows the behavior of these trajectories near the attractor point at large $\xi$ and small $r$. The points on the trajectories shown at the  right panel  correspond to $ \log \xi= -1, 0 , 1$, from top down.  }
\label{xi}
\end{figure}

Cosmological attractors may also appear in the theories describing  non-minimal coupling of scalar fields to gravity \cite{Kallosh:2013tua} of the form
\be
{ {\cal L}_{\rm J}\over \sqrt{-g}}=  {1\over 2} \big (1+ \xi f(\phi) \big )R -{1\over 2} (\partial \phi)^2 - \lambda^2 f^2(\phi)  \ .
\label{ximodels}\ee
where $f(\phi)$ is an arbitrary function. In the particular case $V(\phi) = \lambda^2 f^2(\phi) = \lambda^2\phi^{4}$, these models coincide with the Higgs inflation model  \cite{Salopek:1988qh,Bezrukov:2007ep}. Examples of these inflationary models with  $V(\phi) \sim \phi^{8}, \phi^{6}, \phi^4, \phi^2, \phi, \phi^{2/3}$ were studied in \cite{Kallosh:2013tua} and the $n_s-r$ plots were given, see Fig. \ref{xi}. The plots start at $\xi=0$,  were there is no non-minimal coupling, and then all models are pushed to smaller $r$ with increasing positive $\xi$.  At $\xi \rightarrow \infty$ all models reach the attractor point where $r\approx 3\times 10^{-3}$, as in the   Starobinsky model.

\noindent 
A comparison between Fig.  \ref{xi} for $\xi$-attractors and the closely related Fig. \ref{Fan} for the $\alpha$-attractors reveals  important similarities and differences. In both cases, the attractor mechanism ``saves'' the monomial models, making them compatible with the data. But this happens differently for the $\alpha$-attractors and the $\xi$-attractors. 

The $\xi$-attractor trajectories in the left panel first go down as straight lines parallel to each other, but then they move to the attractor point almost horizontally, spanning large range of values of $n_{s}$ from $0.96$ to $0.98$ for $r \lesssim  0.01$. This makes such models more robust with respect to   future precision data on $n_{s}$. 

On the other hand, the values of $r$ for $\xi$-attractors  do not go much below $3\times 10^{-3}$, which corresponds to the attractor point for $r$ in the limit $\xi \to \infty$. This is a crucial difference as compared to $\alpha$-attractors, which can describe small $r$ all the way down to $r = 0$,  corresponding to the attractor point in the limit $\alpha \to 0$.  

Thus, if gravitational waves with $r \gtrsim 3\times 10^{-3}$ are not found, it  would  disfavor $\xi$-attractors, but such result would be quite compatible with $\alpha$-attractors.
This particular limitation of $\xi$-attractors disappears if one  considers a more general class of models with nonminimal coupling of scalars to gravity 
\be
{ {\cal L}_{\rm J} \over \sqrt{-g}} = \frac{1}{2} \Omega (\phi) R - \frac{1}{2} K_{\rm J} (\phi) (\partial \phi )^2  -V_{\rm J}(\phi)  \, .
\label{lagJ}
\ee  
One can show that  for certain relations between $\Omega (\phi)$, $K_{\rm J} (\phi)$ and $V_{\rm J}(\phi)$ this theory in the Einstein
frame becomes equivalent to the theory of $\alpha$-attractors   \cite{Galante:2014ifa}. Therefore in this more general context one can  describe any small values of $r$.

\

\section{Special cases}\label{spec}

So far we presented T- and E-models with a continuous value of $\alpha$,  which at small $\alpha$ reach the attractor point with cosmological predictions depending on the number of e-foldings and $\alpha$ as shown in \rf{pred}. One can implement these models in the minimal ${\cal N}=1$ supergravity, where the parameter $3\alpha$  is given by
 $3\alpha={1\over 2} |{\cal R}_K|$. Here $|{\cal R}_K|$ is the curvature of  \K\, geometry \cite{Ferrara:2013rsa}.
In the context of the Poincar\'e hyperbolic disk geometry, representing an Escher disk,
$R^2_{\rm Esher} =3\alpha$  defines the size of the disk \cite{Kallosh:2015zsa}.

\vskip  5pt

\begin{figure}[H]
\centering
\includegraphics[scale=0.6]{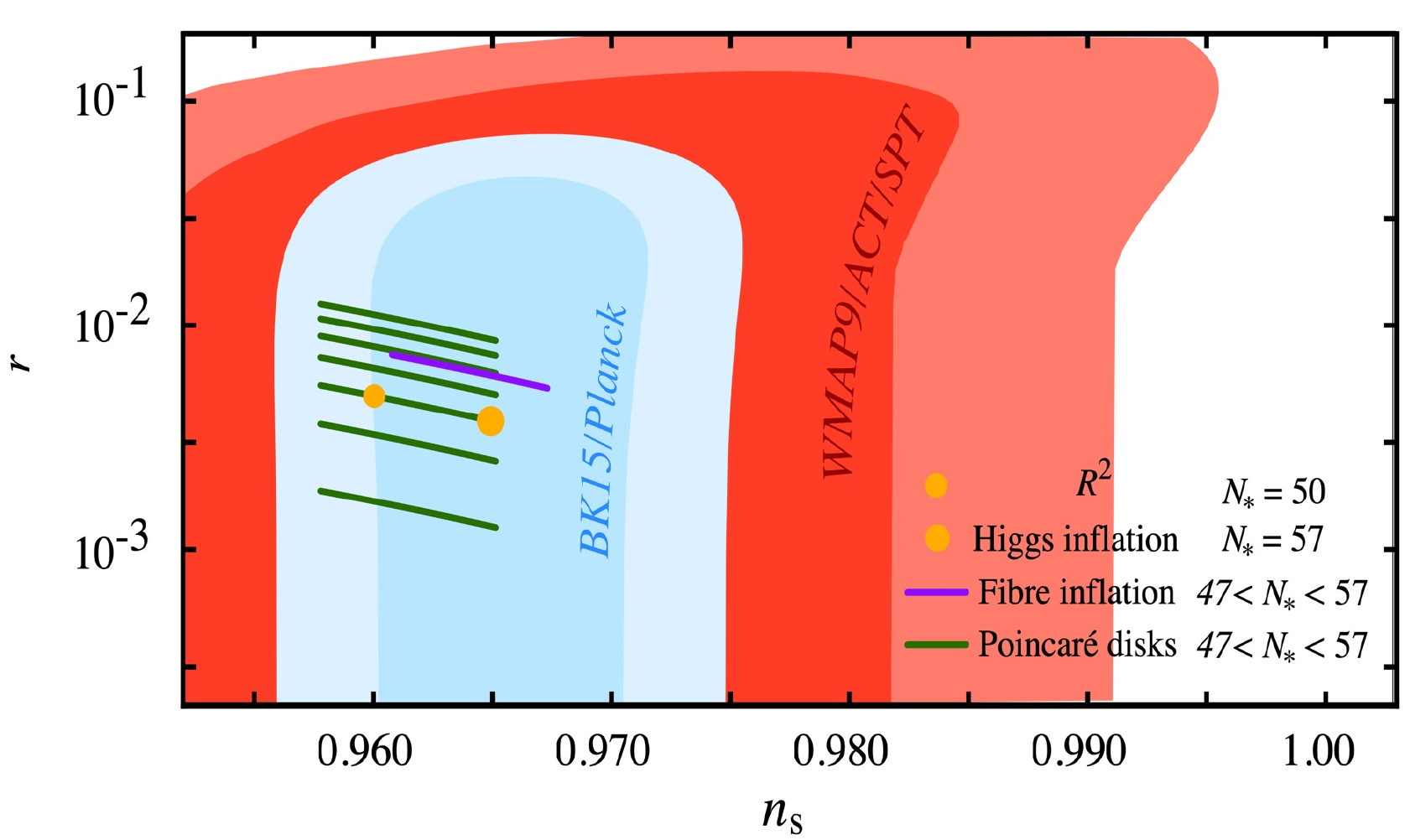}
\caption{\footnotesize This  figure (courtesy of R. Flauger) shows the 7 Poincar\'e disks of the T-model of $\alpha$-attractors as green lines, as well as Higgs inflation, $R^2$ inflation and fibre inflation \cite{Cicoli:2008gp}. The predictions are for $47 < N_e< 57$.}
\label{Flauger}
\end{figure}

The most interesting B-mode targets in this class of cosmological attractor models are the ones with the discrete values of $3\alpha= 7,6,5,4,3,2,1$ 
\cite{Ferrara:2016fwe,Kallosh:2017ced,Gunaydin:2020ric,Kallosh:2021vcf}. These models of Poincar\'e disks are inspired by string theory, M-theory and maximal supergravity. They are known in cosmology community, see Fig. \ref{Flauger}, which shows the plot of R. Flauger presented in his talk at CMB-S4 collaboration meeting in 2021.  

\begin{figure}[!h]
\begin{center}
 \hspace{-3mm}
 \includegraphics[scale=0.32]{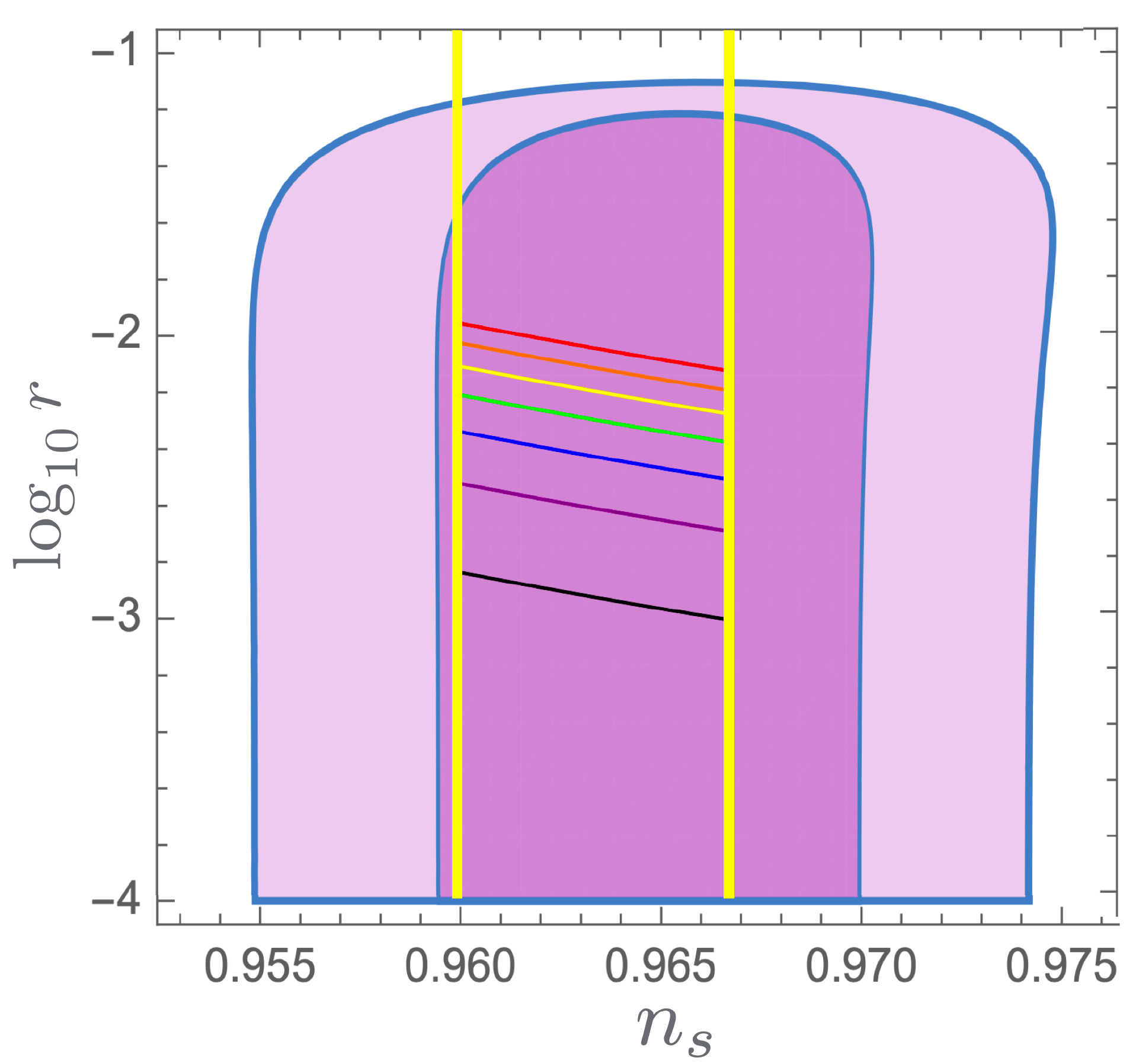}  \hskip 30pt
 \includegraphics[scale=0.326]{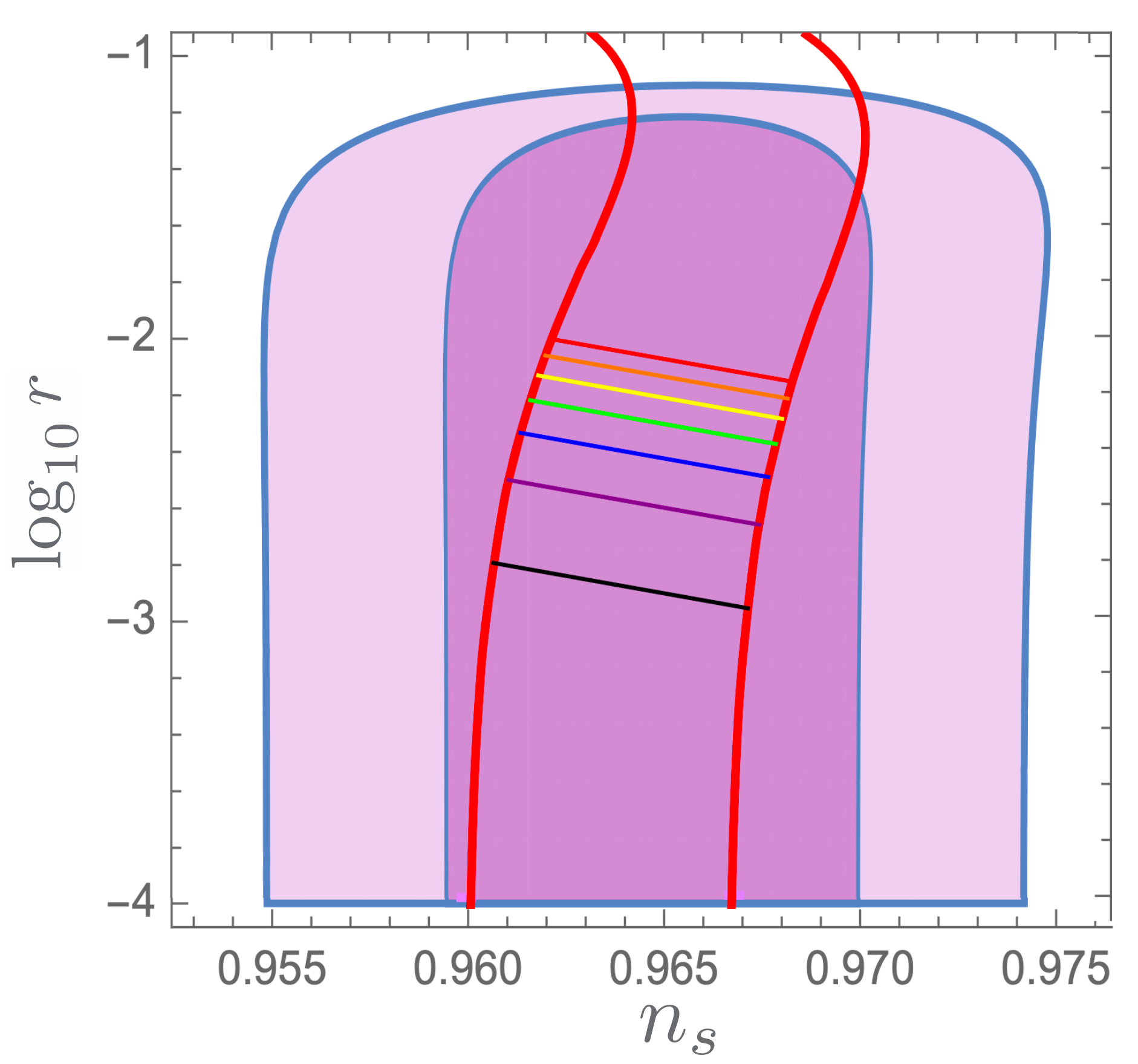}
\end{center}
\vspace{-.7cm}
\caption{\footnotesize
  $\alpha$-attractor benchmarks for T-models (left panel) and E-models (right panel)  show the discrete values of $3\alpha= 7,6,5,4,3,2,1$ from the top going down \cite{Kallosh:2019hzo}.  Dark pink area corresponds to $n_{s}$ and $r$ favored by Planck2018 after taking into account all CMB-related data. The predictions are for $50 < N_e< 60$. }
\label{7disk2}
\end{figure}

  Fig. \ref{7disk2} shows  more detailed plots for the 7 disk  predictions for T- and E-models  \cite{Kallosh:2019hzo}. These predictions correspond to the most  interesting range $10^{{-2}}\lesssim r \lesssim 10^{{-3}}$. 

\begin{figure}[H]
 \vspace{-5mm}
 \centering
\includegraphics[scale=0.45]{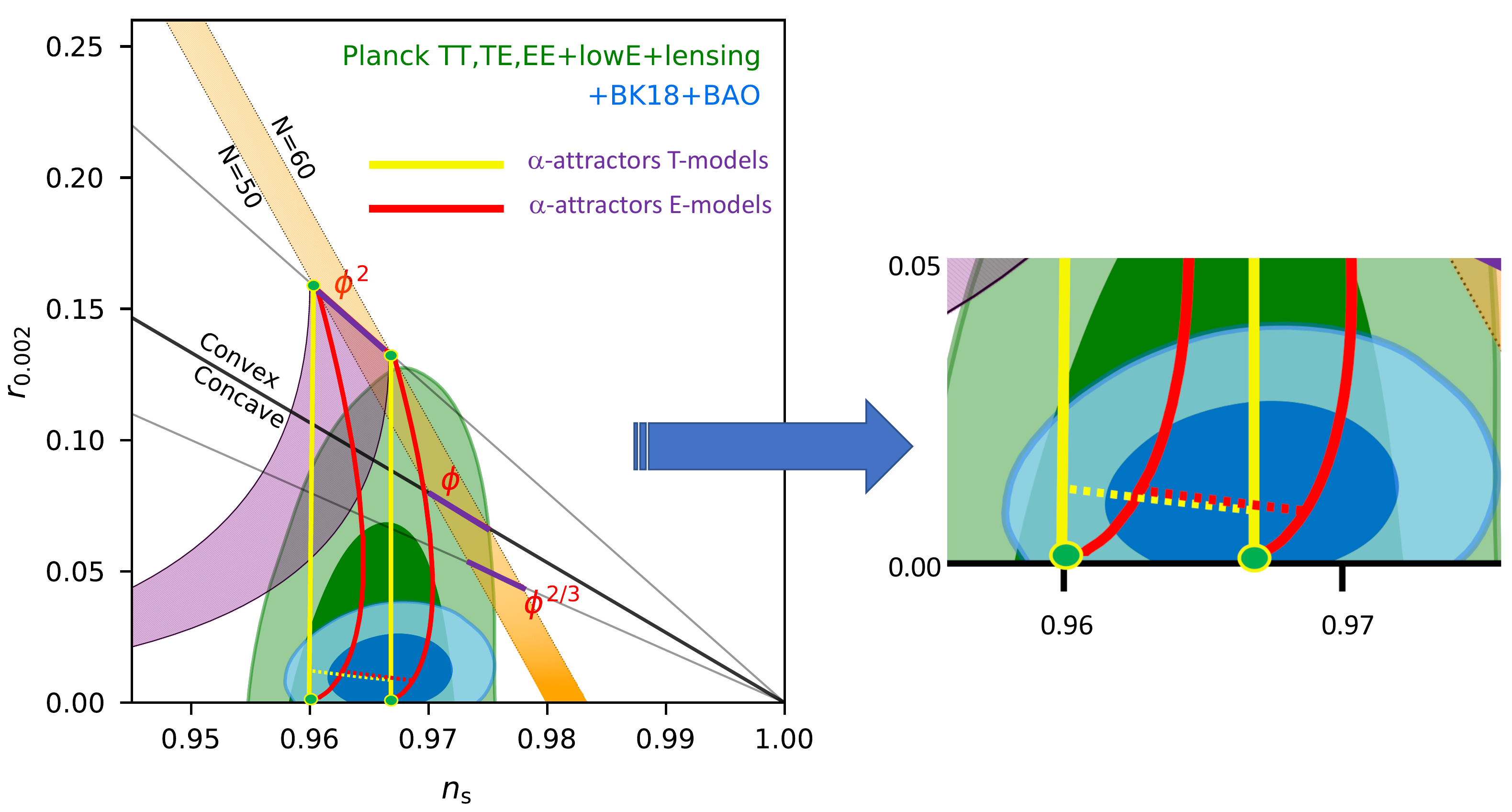}
\caption{\footnotesize  BICEP/Keck results and the predictions for the simplest T-model (yellow dashed line) and E-model (red dashed line)  for  $3\alpha= 7$. These dashed lines   go  through the center of the dark blue area favored by the combination of the Planck, BICEP and Keck results. 
}
\label{7alpha}
\end{figure}

The upper B-mode target with  $3\alpha= 7$, $r \sim 0.01$ is very close to the range that can be explored by BICEP/Keck, if not now, then within the next five years, when the authors of  \cite{BICEPKeck:2021gln} hope to  reach accuracy $\sigma(r) \sim 0.003$.  To illustrate what this might entail, we add to Fig. \ref{TandE} two dashed lines, which show predictions for $3\alpha= 7$ for the simplest T-models (yellow dashed line) and E-model (red dashed line) \cite{Kallosh:2021vcf}. As one can see, these predictions are positioned right at the center of the dark blue ellipse  in Fig. \ref{7alpha}.  

\section{Discussion}

The new BICEP/Keck constraints on the tensor to scalar ratio $r$ strongly disfavor several popular inflationary models, such as natural inflation, the models with monomial potentials, and the Coleman-Weinberg potentials. However, some of these models have powerful theoretical motivation and can have interesting generalizations. For example, the authors of natural inflation proposed the natural chain inflation scenario \cite{Freese:2021noj} which may be compatible with the data. The simplest models of axion monodromy scenario \cite{Silverstein:2008sg,McAllister:2014mpa} lead to monomial potentials, but allow for various modifications changing the predicted values of $n_{s}$ and $r$, see e.g.  \cite{Wenren:2014cga,DAmico:2021vka}.

A particularly  interesting inflationary model, which fit the Planck/BICEP/Keck data, is the fibre inflation model based on string theory  \cite{Cicoli:2008gp} with the prediction $r \sim 0.007$ indicated by a purple line in Fig. \ref{Flauger}. Other examples of  inflationary models  which can be compatible with the current and future data  can be found, in particular, in \cite{Enckell:2018uic,Ellis:2020lnc,Hazra:2021eqk}. 

It is most interesting that some models proposed many decades ago and based on entirely different ideas, such as the Starobinsky model  \cite{Starobinsky:1980te}, the Higgs inflation model  \cite{Salopek:1988qh,Bezrukov:2007ep}, and the GL model  \cite{Goncharov:1983mw,Linde:2014hfa}, require just  a single parameter to successfully account for all presently available data.   In this paper we described a broad class of  cosmological attractors \cite{Kallosh:2013hoa,Ferrara:2013rsa,Kallosh:2013yoa,Galante:2014ifa,Kallosh:2015zsa,Kallosh:2019eeu,Kallosh:2019hzo}, which generalized the three models mentioned above.  

As discussed in Section \ref{sattr}, $\alpha$-attractors, such as T-models  \rf{cosmoqq} and E-models   \rf{apole},
provide  a good fit to the Planck and BICEP/Keck data and   have ample   flexibility to describe  {\it any}\, value of $r$  below the BICEP/Keck bound $0.036$.   A broad class of pole inflation models \rf{actionQ}, D-brane models  \rf{KKLTI},  and models describing general  non-minimal coupling of scalar fields to gravity  \rf{lagJ}  also can describe inflation at very small  $r$,   all the way down to $r = 0$.

On the other hand, some string theory inspired versions of $\alpha$-attractors described in Section \ref{spec} predict a discrete spectrum of 7 different  values of r in the range from $10^{-2}$ to $10^{{-3}}$. The upper one of these predictions is shown in Fig. \ref{7alpha} by the dashed lines going through the center of the dark blue area favored by the combination of the Planck, BICEP and Keck results. 

At present, the error bars of the BICEP/Keck estimate $r = 0.014 \pm 0.01$ are large, $\sigma(r) \sim 0.009$. However, the authors of  \cite{BICEPKeck:2021gln} expect that within the next few years they may improve the accuracy up to $\sigma(r) \sim 0.003$.  This suggests that the model describing the first discrete  target $r\approx 10^{-2}$ for $\alpha$-attractors  with $3\alpha=7$ may be either confirmed or ruled out. It will take much more time and probably a satellite mission to reach  $r\approx 10^{-3}$, corresponding to the last Poincar\'e disk in  Figs. \ref{Flauger}, \ref{7disk2}  with  $3\alpha=1$.

\vskip 3pt

%\subsection*{Comparison with new Planck/BICEP/Keck constraints  \cite{Tristram:2021tvh}.}

\begin{figure}[H]
\centering
\includegraphics[scale=0.46]{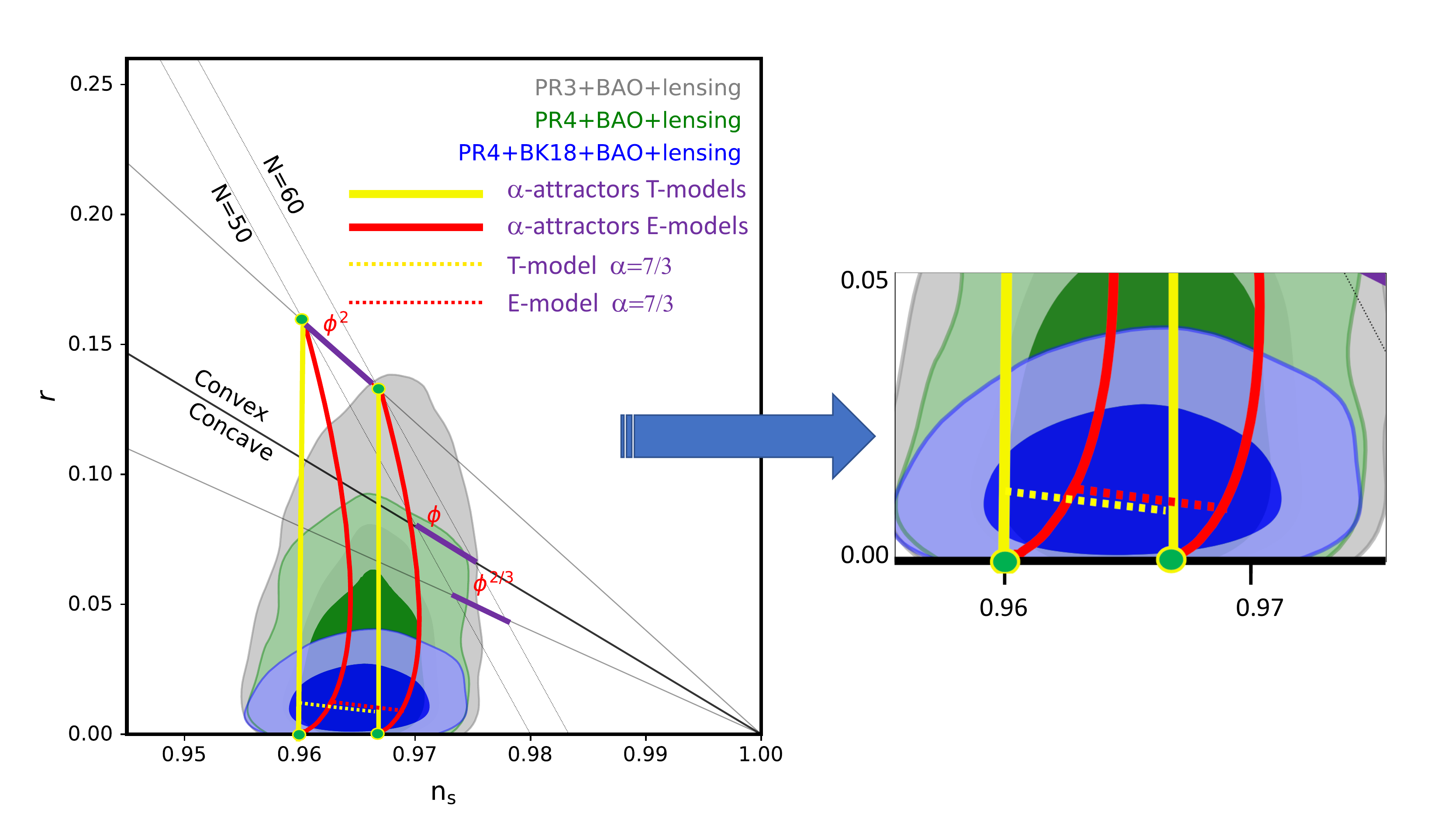}
\caption{\footnotesize  Results of \cite{Tristram:2021tvh} and the predictions for the simplest T-model (yellow dashed line) and E-model (red dashed line)  for  $3\alpha= 7$. These dashed lines  go  through the center of the dark blue area favored by the combination of the Planck, BICEP and Keck results according to \cite{Tristram:2021tvh}. 
}
\label{TR}
\end{figure}

{\bf Addendum:} After this paper was submitted, a new set of constraints on $r$ and $n_{s}$ was given in \cite{Tristram:2021tvh} after a somewhat different analysis of Planck data.  The  main results of \cite{Tristram:2021tvh} are very similar to those reported in  \cite{BICEPKeck:2021gln}. According to  \cite{Tristram:2021tvh}, the 2$\sigma$ upper bound on $r$ changes from $r < 0.036$ \cite{BICEPKeck:2021gln} to $r < 0.032$. The  1$\sigma$ bound on $n_{s}$ presented in  \cite{Tristram:2021tvh} is shifted to smaller values of $n_{s} $ by  $\Delta n_{s} \sim -0.003$. This does  not affect conclusions of  our paper. 
If anything, the new constraints given in  \cite{Tristram:2021tvh} make  the match between the observations and the predictions of cosmological attractors  even better. One can see it by comparing Fig. \ref{7alpha} with  Fig. \ref{TR}, which shows the results of  \cite{Tristram:2021tvh} and the predictions of the simplest T- and E-models of $\alpha$-attractors. 

\section*{Acknowledgement}
We are grateful to  G. Efstathiou, S. Ferrara,  R. Flauger,  N. Kaloper, C. L. Kuo, D. Roest, T. Wrase and Y. Yamada  for stimulating discussions.  This work is  supported by SITP and by the US National Science Foundation Grant  PHY-2014215, and by the  Simons Foundation Origins of the Universe program (Modern Inflationary Cosmology collaboration).

\appendix

\bibliographystyle{JHEP}
\bibliography{lindekalloshrefs}

\providecommand{\href}[2]{#2}\begingroup\raggedright\begin{thebibliography}{10}

\bibitem{BICEPKeck:2021gln}
{\scshape BICEP/Keck} collaboration, \emph{{Improved Constraints on Primordial
  Gravitational Waves using Planck, WMAP, and BICEP/Keck Observations through
  the 2018 Observing Season}},
  \href{https://doi.org/10.1103/PhysRevLett.127.151301}{\emph{Phys. Rev. Lett.}
  {\bfseries 127} (2021) 151301}
  [\href{https://arxiv.org/abs/2110.00483}{{\ttfamily 2110.00483}}].

\bibitem{Freese:1990rb}
K.~Freese, J.A.~Frieman and A.V.~Olinto, \emph{{Natural inflation with pseudo -
  Nambu-Goldstone bosons}},
  \href{https://doi.org/10.1103/PhysRevLett.65.3233}{\emph{Phys. Rev. Lett.}
  {\bfseries 65} (1990) 3233}.

\bibitem{Planck:2018jri}
{\scshape Planck} collaboration, \emph{{Planck 2018 results. X. Constraints on
  inflation}}, \href{https://doi.org/10.1051/0004-6361/201833887}{\emph{Astron.
  Astrophys.} {\bfseries 641} (2020) A10}
  [\href{https://arxiv.org/abs/1807.06211}{{\ttfamily 1807.06211}}].

\bibitem{Kallosh:2019jnl}
R.~Kallosh and A.~Linde, \emph{{On hilltop and brane inflation after Planck}},
  \href{https://doi.org/10.1088/1475-7516/2019/09/030}{\emph{JCAP} {\bfseries
  09} (2019) 030} [\href{https://arxiv.org/abs/1906.02156}{{\ttfamily
  1906.02156}}].

\bibitem{Linde:1981mu}
A.D.~Linde, \emph{{A New Inflationary Universe Scenario: A Possible Solution of
  the Horizon, Flatness, Homogeneity, Isotropy and Primordial Monopole
  Problems}}, \href{https://doi.org/10.1016/0370-2693(82)91219-9}{\emph{Phys.
  Lett.} {\bfseries B108} (1982) 389}.

\bibitem{Albrecht:1982wi}
A.~Albrecht and P.J.~Steinhardt, \emph{{Cosmology for Grand Unified Theories
  with Radiatively Induced Symmetry Breaking}},
  \href{https://doi.org/10.1103/PhysRevLett.48.1220}{\emph{Phys. Rev. Lett.}
  {\bfseries 48} (1982) 1220}.

\bibitem{Kallosh:2013hoa}
R.~Kallosh and A.~Linde, \emph{{Universality Class in Conformal Inflation}},
  \href{https://doi.org/10.1088/1475-7516/2013/07/002}{\emph{JCAP} {\bfseries
  1307} (2013) 002} [\href{https://arxiv.org/abs/1306.5220}{{\ttfamily
  1306.5220}}].

\bibitem{Ferrara:2013rsa}
S.~Ferrara, R.~Kallosh, A.~Linde and M.~Porrati, \emph{{Minimal Supergravity
  Models of Inflation}},
  \href{https://doi.org/10.1103/PhysRevD.88.085038}{\emph{Phys. Rev.}
  {\bfseries D88} (2013) 085038}
  [\href{https://arxiv.org/abs/1307.7696}{{\ttfamily 1307.7696}}].

\bibitem{Kallosh:2013yoa}
R.~Kallosh, A.~Linde and D.~Roest, \emph{{Superconformal Inflationary
  $\alpha$-Attractors}},
  \href{https://doi.org/10.1007/JHEP11(2013)198}{\emph{JHEP} {\bfseries 11}
  (2013) 198} [\href{https://arxiv.org/abs/1311.0472}{{\ttfamily 1311.0472}}].

\bibitem{Galante:2014ifa}
M.~Galante, R.~Kallosh, A.~Linde and D.~Roest, \emph{{Unity of Cosmological
  Inflation Attractors}},
  \href{https://doi.org/10.1103/PhysRevLett.114.141302}{\emph{Phys. Rev. Lett.}
  {\bfseries 114} (2015) 141302}
  [\href{https://arxiv.org/abs/1412.3797}{{\ttfamily 1412.3797}}].

\bibitem{Kallosh:2015zsa}
R.~Kallosh and A.~Linde, \emph{{Escher in the Sky}},
  \href{https://doi.org/10.1016/j.crhy.2015.07.004}{\emph{Comptes Rendus
  Physique} {\bfseries 16} (2015) 914}
  [\href{https://arxiv.org/abs/1503.06785}{{\ttfamily 1503.06785}}].

\bibitem{Kallosh:2019eeu}
R.~Kallosh and A.~Linde, \emph{{B-mode Targets}},
  \href{https://doi.org/10.1016/j.physletb.2019.134970}{\emph{Phys. Lett. B}
  {\bfseries 798} (2019) 134970}
  [\href{https://arxiv.org/abs/1906.04729}{{\ttfamily 1906.04729}}].

\bibitem{Kallosh:2019hzo}
R.~Kallosh and A.~Linde, \emph{{CMB Targets after PlanckCMB targets after the
  latest $Planck$ data release}},
  \href{https://doi.org/10.1103/PhysRevD.100.123523}{\emph{Phys. Rev.}
  {\bfseries D100} (2019) 123523}
  [\href{https://arxiv.org/abs/1909.04687}{{\ttfamily 1909.04687}}].

\bibitem{Kallosh:2016gqp}
R.~Kallosh and A.~Linde, \emph{{Cosmological Attractors and Asymptotic Freedom
  of the Inflaton Field}},
  \href{https://doi.org/10.1088/1475-7516/2016/06/047}{\emph{JCAP} {\bfseries
  1606} (2016) 047} [\href{https://arxiv.org/abs/1604.00444}{{\ttfamily
  1604.00444}}].

\bibitem{Starobinsky:1980te}
A.A.~Starobinsky, \emph{{A New Type of Isotropic Cosmological Models Without
  Singularity}},
  \href{https://doi.org/10.1016/0370-2693(80)90670-X}{\emph{Phys. Lett.}
  {\bfseries 91B} (1980) 99}.

\bibitem{Kallosh:2018zsi}
R.~Kallosh, A.~Linde and Y.~Yamada, \emph{{Planck 2018 and Brane Inflation
  Revisited}}, \href{https://doi.org/10.1007/JHEP01(2019)008}{\emph{JHEP}
  {\bfseries 01} (2019) 008}
  [\href{https://arxiv.org/abs/1811.01023}{{\ttfamily 1811.01023}}].

\bibitem{Dvali:1998pa}
G.R.~Dvali and S.H.H.~Tye, \emph{{Brane inflation}},
  \href{https://doi.org/10.1016/S0370-2693(99)00132-X}{\emph{Phys. Lett. B}
  {\bfseries 450} (1999) 72}
  [\href{https://arxiv.org/abs/hep-ph/9812483}{{\ttfamily hep-ph/9812483}}].

\bibitem{Kachru:2003sx}
S.~Kachru, R.~Kallosh, A.D.~Linde, J.M.~Maldacena, L.P.~McAllister and
  S.P.~Trivedi, \emph{{Towards inflation in string theory}},
  \href{https://doi.org/10.1088/1475-7516/2003/10/013}{\emph{JCAP} {\bfseries
  0310} (2003) 013} [\href{https://arxiv.org/abs/hep-th/0308055}{{\ttfamily
  hep-th/0308055}}].

\bibitem{Dong:2010in}
X.~Dong, B.~Horn, E.~Silverstein and A.~Westphal, \emph{{Simple exercises to
  flatten your potential}},
  \href{https://doi.org/10.1103/PhysRevD.84.026011}{\emph{Phys. Rev.}
  {\bfseries D84} (2011) 026011}
  [\href{https://arxiv.org/abs/1011.4521}{{\ttfamily 1011.4521}}].

\bibitem{DAmico:2017cda}
G.~D'Amico, N.~Kaloper and A.~Lawrence, \emph{{Monodromy Inflation in the
  Strong Coupling Regime of the Effective Field Theory}},
  \href{https://doi.org/10.1103/PhysRevLett.121.091301}{\emph{Phys. Rev. Lett.}
  {\bfseries 121} (2018) 091301}
  [\href{https://arxiv.org/abs/1709.07014}{{\ttfamily 1709.07014}}].

\bibitem{Kallosh:2013tua}
R.~Kallosh, A.~Linde and D.~Roest, \emph{{Universal Attractor for Inflation at
  Strong Coupling}},
  \href{https://doi.org/10.1103/PhysRevLett.112.011303}{\emph{Phys. Rev. Lett.}
  {\bfseries 112} (2014) 011303}
  [\href{https://arxiv.org/abs/1310.3950}{{\ttfamily 1310.3950}}].

\bibitem{Salopek:1988qh}
D.S.~Salopek, J.R.~Bond and J.M.~Bardeen, \emph{{Designing Density Fluctuation
  Spectra in Inflation}},
  \href{https://doi.org/10.1103/PhysRevD.40.1753}{\emph{Phys. Rev.} {\bfseries
  D40} (1989) 1753}.

\bibitem{Bezrukov:2007ep}
F.L.~Bezrukov and M.~Shaposhnikov, \emph{{The Standard Model Higgs boson as the
  inflaton}}, \href{https://doi.org/10.1016/j.physletb.2007.11.072}{\emph{Phys.
  Lett.} {\bfseries B659} (2008) 703}
  [\href{https://arxiv.org/abs/0710.3755}{{\ttfamily 0710.3755}}].

\bibitem{Ferrara:2016fwe}
S.~Ferrara and R.~Kallosh, \emph{{Seven-disk manifold, $\alpha$-attractors, and
  $B$ modes}}, \href{https://doi.org/10.1103/PhysRevD.94.126015}{\emph{Phys.
  Rev.} {\bfseries D94} (2016) 126015}
  [\href{https://arxiv.org/abs/1610.04163}{{\ttfamily 1610.04163}}].

\bibitem{Kallosh:2017ced}
R.~Kallosh, A.~Linde, T.~Wrase and Y.~Yamada, \emph{{Maximal Supersymmetry and
  B-Mode Targets}}, \href{https://doi.org/10.1007/JHEP04(2017)144}{\emph{JHEP}
  {\bfseries 04} (2017) 144}
  [\href{https://arxiv.org/abs/1704.04829}{{\ttfamily 1704.04829}}].

\bibitem{Gunaydin:2020ric}
M.~Gunaydin, R.~Kallosh, A.~Linde and Y.~Yamada, \emph{{M-theory Cosmology,
  Octonions, Error Correcting Codes}},
  \href{https://doi.org/10.1007/JHEP01(2021)160}{\emph{JHEP} {\bfseries 01}
  (2021) 160} [\href{https://arxiv.org/abs/2008.01494}{{\ttfamily
  2008.01494}}].

\bibitem{Kallosh:2021vcf}
R.~Kallosh, A.~Linde, T.~Wrase and Y.~Yamada, \emph{{IIB String Theory and
  Sequestered Inflation}},  \href{https://arxiv.org/abs/2108.08492}{{\ttfamily
  2108.08492}}.

\bibitem{Cicoli:2008gp}
M.~Cicoli, C.P.~Burgess and F.~Quevedo, \emph{{Fibre Inflation: Observable
  Gravity Waves from IIB String Compactifications}},
  \href{https://doi.org/10.1088/1475-7516/2009/03/013}{\emph{JCAP} {\bfseries
  0903} (2009) 013} [\href{https://arxiv.org/abs/0808.0691}{{\ttfamily
  0808.0691}}].

\bibitem{Freese:2021noj}
K.~Freese, A.~Litsa and M.W.~Winkler, \emph{{Natural Chain Inflation}},
  \href{https://arxiv.org/abs/2109.11556}{{\ttfamily 2109.11556}}.

\bibitem{Silverstein:2008sg}
E.~Silverstein and A.~Westphal, \emph{{Monodromy in the CMB: Gravity Waves and
  String Inflation}},
  \href{https://doi.org/10.1103/PhysRevD.78.106003}{\emph{Phys. Rev.}
  {\bfseries D78} (2008) 106003}
  [\href{https://arxiv.org/abs/0803.3085}{{\ttfamily 0803.3085}}].

\bibitem{McAllister:2014mpa}
L.~McAllister, E.~Silverstein, A.~Westphal and T.~Wrase, \emph{{The Powers of
  Monodromy}}, \href{https://doi.org/10.1007/JHEP09(2014)123}{\emph{JHEP}
  {\bfseries 09} (2014) 123} [\href{https://arxiv.org/abs/1405.3652}{{\ttfamily
  1405.3652}}].

\bibitem{Wenren:2014cga}
D.~Wenren, \emph{{Tilt and Tensor-to-Scalar Ratio in Multifield Monodromy
  Inflation}},  \href{https://arxiv.org/abs/1405.1411}{{\ttfamily 1405.1411}}.

\bibitem{DAmico:2021vka}
G.~D'Amico, N.~Kaloper and A.~Westphal, \emph{{Double Monodromy Inflation: A
  Gravity Waves Factory for CMB-S4, LiteBIRD and LISA}},
  \href{https://arxiv.org/abs/2101.05861}{{\ttfamily 2101.05861}}.

\bibitem{Enckell:2018uic}
V.-M.~Enckell, K.~Enqvist, S.~Rasanen and L.-P.~Wahlman, \emph{{Higgs-$R^2$
  inflation - full slow-roll study at tree-level}},
  \href{https://doi.org/10.1088/1475-7516/2020/01/041}{\emph{JCAP} {\bfseries
  01} (2020) 041} [\href{https://arxiv.org/abs/1812.08754}{{\ttfamily
  1812.08754}}].

\bibitem{Ellis:2020lnc}
J.~Ellis, M.A.G.~Garcia, N.~Nagata, N.D.~V., K.A.~Olive and S.~Verner,
  \emph{{Building models of inflation in no-scale supergravity}},
  \href{https://doi.org/10.1142/S0218271820300116}{\emph{Int. J. Mod. Phys. D}
  {\bfseries 29} (2020) 2030011}
  [\href{https://arxiv.org/abs/2009.01709}{{\ttfamily 2009.01709}}].

\bibitem{Hazra:2021eqk}
D.K.~Hazra, D.~Paoletti, I.~Debono, A.~Shafieloo, G.F.~Smoot and
  A.A.~Starobinsky, \emph{{Inflation Story: slow-roll and beyond}},
  \href{https://arxiv.org/abs/2107.09460}{{\ttfamily 2107.09460}}.

\bibitem{Goncharov:1983mw}
A.B.~Goncharov and A.D.~Linde, \emph{{Chaotic Inflation in Supergravity}},
  \href{https://doi.org/10.1016/0370-2693(84)90027-3}{\emph{Phys. Lett.}
  {\bfseries B139} (1984) 27}.

\bibitem{Linde:2014hfa}
A.~Linde, \emph{{Does the first chaotic inflation model in supergravity provide
  the best fit to the Planck data?}},
  \href{https://doi.org/10.1088/1475-7516/2015/02/030}{\emph{JCAP} {\bfseries
  1502} (2015) 030} [\href{https://arxiv.org/abs/1412.7111}{{\ttfamily
  1412.7111}}].

\bibitem{Tristram:2021tvh}
M.~Tristram et~al., \emph{{Improved limits on the tensor-to-scalar ratio using
  BICEP and Planck}},  \href{https://arxiv.org/abs/2112.07961}{{\ttfamily
  2112.07961}}.

\end{thebibliography}\endgroup
\end{document}